\documentclass[12pt]{article}

\usepackage{epsf}
\usepackage[dvips]{graphicx}

\begin{document}

\begin{minipage}{14cm}
\vskip 2cm
\end{minipage}
\vskip 2cm

\begin{center}
{\bfseries JET ENERGY DENSITY IN HADRON-HADRON COLLISIONS\\
 AT HIGH ENERGIES}

\vskip 5mm

J.V. Ili\' c$^{1}$, G.P. \v {S}koro$^{2 \dag}$ and M.V.
Tokarev$^{3 \star}$

\vskip 5mm

{\small (1) {\it Institute of Nuclear Sciences "Vin\v {c}a",
Belgrade, Serbia and Montenegro }
\\
(2) {\it Faculty of Physics, University of Belgrade, Belgrade,
Serbia and Montenegro }
\\
(3) {\it Veksler and Baldin Laboratory of High Energies, Joint
Institute for Nuclear Research, Dubna, Russia}
\\
$^{\dag}${\it E-mail: goran@ff.bg.ac.yu }\\
$^{\star}${\it E-mail: tokarev@sunhe.jinr.ru} }
\end{center}

\vskip 20mm

\begin{center}
\begin{minipage}{150mm}
\centerline{\bf Abstract}
The average particle multiplicity
density $dN/d\eta$ is the dynamical quantity which reflects some
regularities of particle production in low-$p_T$ range. The
quantity is an important ingredient of $z$-scaling.
Experimental results on charged particle density are available for
$pp$, $pA$ and $AA$ collisions while experimental properties of
the jet density are still an open question. The goal of this work
is to find the variable which will reflect the main features of
the jet production in low transverse energy range and play the
role of the scale factor for the scaling function $\psi(z)$ and
variable $z$ in data $z$-presentation. The appropriate  candidate is the
variable we called "scaled jet energy density". Scaled jet energy
density is the probability to have a jet with defined $E_T$ in
defined $x_T$ and pseudorapidity regions. The PYTHIA6.2 Monte
Carlo generator is used for calculation of scaled jet energy
density in proton-proton collisions over a high energy range
($\sqrt s = 200-14000$~GeV) and at $\eta = 0$. The properties of
the new variable are discussed and sensitivity to "physical scenarios"
applied in the standard Monte Carlo generator is noted.
The results of scaled jet energy density at LHC energies
are presented and compared with predictions based on $z$-scaling.

\end{minipage}
\end{center}

\vspace*{2cm}
\begin{center}
{\it
    Submitted to "Physics of Particles and Nuclei, Letters"
}
\end{center}

\newpage

\vskip 5mm

{\section{Introduction}}
For the description of particle production
in high-$p_T$ $pp$, $\bar pp$ and $pA$ collisions at high energies,
the $z$-scaling concept is proposed in \cite{z1}.
In the framework of $z$-scaling, such experimental observables as inclusive cross-section
$Ed^3\sigma/dp^3$ and
the average charged particle multiplicity density
$\rho\equiv dN/d\eta$
are used to construct the scaling function
 $\psi(z)$ and variable $z$.
The scaling, known as $z$-scaling, reveals interesting properties. There are the
independence of the scaling function, $\psi(z)$, on collision energy and an angle of
produced objects (hadron, photon). A general concept of the scaling is based on such
fundamental principles as self-similarity, locality, fractality
and scale-relativity \cite{z2, Not}.
Because the scaling function $\psi(z)$ is well defined in  hadron-hadron
collisions and  expressed via two experimental observables,
it is clear that the quantity can be used to study the properties of jet production, too.

In $z$-scaling concept, the average charged particle multiplicity density
plays the role of the scale factor, $z\sim 1/\rho(s)$, and
 $ \psi(z) \sim 1/\rho(s,\eta)$.
 Experimental results on charged
particle density are available for $pp$, $pA$ and $AA$ collisions
while experimental properties of the jet density are still an open
question. In the case of jets, there are a lot of uncertainties
(knowledge of parton distribution and fragmentation functions, knowledge of
factorization, renormalization and fragmentation scales,
uncertainties in the parton shower modelling etc.,)
causing the problems in understanding of jet behavior at very high energies.
The goal of this work is to find the variable which will reflects
the main features of the jet production in low transverse energy range at a given energy
and play the role of the scale factor.

The paper is organized as follows. A basic description of a scale
factor in $z$-scaling concept as well as results of
Monte Carlo simulations on a scale factor in the case of charged
particles production are given in Sec.2.
New results on a scale factor for jet production based
on the analysis of the experimental data and Monte Carlo
simulations are described in Sec.3.
Discussion of the obtained results at the LHC energies
is presented in Sec.4. Conclusions are summarized in Sec.5.

{\section{Scale factor in $z$-scaling concept}}
One of the most interesting problems in the modern particle physics
is a search for general properties of quark and
gluon interactions in collisions of leptons, hadrons and nuclei.
Universal approach to description of the
processes allows us detail understanding of the physical
phenomena underlying the secondary particle production.
Up to date, the investigation of hadron properties in the high energy
collisions has revealed widely known scaling regularities.
Some of the most popular and famous are the Feynman scaling \cite{Feynman}
for inclusive hadron production,
the Bjorken scaling observed in
deep inelastic scattering (DIS) \cite{Bjorken}, y-scaling valid
in DIS on nuclei \cite{Y}, limiting fragmentation established for
nuclei fragmentation \cite{Yang}, scaling behaviour of the
cumulative particle production \cite{Baldin,Stavinsky,Leksin}, KNO
scaling \cite{KNO} and others. However, detailed experimental study of the
established scaling laws has shown certain violations of these.
The domains in which the observed regularities are violated is
of great interest. These can be relevant in searching for
new physical phenomena - quark compositeness, new interactions,
quark-gluon plasma and others.

The concept of the $z$-scaling is introduced in \cite{z1} for the
description of inclusive production cross sections in
$pp/\overline pp$  interactions at high energies and high $p_T$ values of
secondary particles.
The scaling function $\psi (z)$ is expressed via the invariant
inclusive cross section $Ed^3\sigma/dp^3$
and the average charged particle multiplicity density $\rho (s,\eta)$.
The function  $\psi (z)$ is found to be independent
of collision energy ${\sqrt s}$ and an angle ${\theta}$ of the inclusive
particle.
The scaling was also applied for the analysis of the inclusive
particle productions in $pA$ collisions \cite{z2}, jet productions \cite{TD}, etc.
The scaling function of direct photon production was found to reveal
the power behavior of $\psi (z)\sim z^{-\beta}$ \cite{Pot}.
The properties of the scaling are assumed to reflect
the fundamental properties of particle structure, interaction and production. The
scaling function describes the probability to form the produced particle with formation
length $z$. The existence of the scaling itself means that the hadronization mechanism
of particle production reveals such fundamental properties as self-similarity,
locality, fractality and scale-relativity.

But, it was also found that there is a strong sensitivity
of the scaling behavior on the energy dependence of the scale factor $\rho(s)$ at $\eta=0$.
The experimental results show that scale factor $\rho(s)$
(the average charged particle multiplicity density)
is well defined quantity
(at least up to Tevatron energies) and that simulation results of standard
Monte Carlo generators (as PYTHIA) are in nice agreement
with available experimental data. But, it is clear that this scale factor
cannot be used for description of processes  in the case of
jet production at high energies and that corresponding variable for
jets must be found. This variable should represent the
main properties of jet production at low $E_T$ and must be, as much as possible,
independent of jet energy $E_T$. It should be noted
that the scale factor $\rho(s,\eta)$ in the case of particle
production has such properties. Because of that and for the sake of completeness,
 we start the story about the jet scale factor with
short description of the properties of charged particle multiplicity density
based on the results of Monte Carlo simulations.

The PYTHIA Monte Carlo generator \cite{PYTHIA} is used for
calculations of charged particle multiplicity density in hadron-hadron
($pp$, ${\pi}p$)
collisions in high energy range and at pseudorapidity $\eta = 0$.
In both the cases, the dependence of density $\rho$  on energy,
$\sqrt s$, at $\eta = 0$ was fitted by the function:
$\rho(s) = a{\cdot}s^{b}$,
where $a$ and $b$ are free parameters.
Choice of the fitting function reflects the experimentally observed
power law dependence of charged particle density on energy.
On the other hand, the properties of this power law should be
a consequence of the Pomeron trajectory
with intercept $\Delta = \alpha_P -1$.
Based on analysis of available experimental data
the value of the quantity was found to be $ 0.105$.
Charged particle density $\rho(s)$ in $pp$ interactions was
simulated in the energy range $\sqrt s = 50 \div 14000$~GeV. The value of
$d{\sigma}^{ch}/d{\eta}$ for every energy was normalized to the
corresponding value of the inelastic cross-section ${\sigma}_{inel}$.
The results of
simulations are shown in Figure 1(a). As can be seen, the fit is
satisfactory, with parameters equal to $a = 0.74(12)$ and
$b=0.105(11)$.
This result fully agrees with theoretical predictions and
available experimental results.
It is expected
that multiplicity density of charged particles at $\sqrt s =14$~TeV will follow the
same energy dependence but it is, in principle, still an open question.
The Monte Carlo
simulations of charged particle density at LHC which are in progress (see, for
example, ATLAS TDR, p.480 \cite {Atlas}) give results for
multiplicity density at $\eta = 0$ in the range from $4.5$ up to $10$.
Charged particle density $dN^{ch}/d\eta$ in ${\pi}p$ collisions was
simulated in the energy range $\sqrt s = 10 \div 200$~GeV. The chosen energy
range is relatively narrow, but it is, at the
moment, experimentally available. The results of
simulations are presented in Figure 1(b). The parameter values were found to be
$a = 0.59(8)$ and $b=0.126(17)$.
For $pp$ and ${\pi}p$ collisions, we have obtained
practically the same value for parameter $b$ (within the errors).
Also, in the energy region from 50 to 200~GeV there is no
difference between densities in $pp$ and ${\pi}p$ collisions.

The power law dependence of charged particle density on energy $\sqrt s$
is valid for $pA$ too.
In the case of $pA$ collisions,
the densities of charged particles can be parameterized \cite{z2} by
the formula: ${dN^{ch}}/{d\eta} \simeq 0.67{\cdot}A^{0.18}{\cdot}s^{0.105}$,
where $A$ is the atomic weight of the corresponding nucleus.

{\section{Jet energy density}}
In the case of jets, the situation is much more complicated.
For example, in \cite{TD}, the average jet multiplicity density
 dependence on energy $\rho (s)$,
resulted from requirements of z-scaling, is used for
analysis of jet production at high energies. The authors used different experimental results on
jet cross-sections to produce semi-empirical energy dependence of  jet scale factor.
The result of that analysis is reproduced in Table 1.
Also, the authors give the prediction of jet multiplicity density at LHC energies but
emphasized that high accuracy measurements of absolute cross section normalization and the jet
density are very important to verify the energy independence of the scaling function $\psi(z)$.
On the other hand, the search for the "universal" jet scale factor is complicated
because the cross sections for jet production have non-trivial behavior.
The cross sections for
production of  jets with a fixed transverse energy $E_T$ rise with $\sqrt s$. This is
because the important $x$ values decrease and there are more partons at smaller $x$.
But, cross sections for jets with transverse momentum that is a fixed fraction of
$\sqrt s$ fall with $\sqrt s$. This is mostly because
the partonic cross sections fall with $E_T$ like ${E_T}^{-2}$.

\vskip 0.7cm
\noindent
{\it{Table 1}. The average jet multiplicity density $\rho (s)$ normalized to the value at
$\sqrt s = 1800~GeV$ in
$\bar pp$ and $pp$ collisions over the
central pseudorapidity range as a function of collision energy \cite{TD}.}

\begin{center}
\vskip 0.2cm
{\baselineskip=1.5\baselineskip
\begin{tabular}{|c|c|c|c|c|c|c|c|}
\hline
{$\sqrt s$ [GeV]} & {63} & {200} & {630} & {1000} & {1800} & {7000} & {14000}\\ \hline
{Average jet density (normalized)} & {0.35} & {0.5} & {0.67} & {0.84} & {1} & {1.57} & {1.95}\\
\hline
\end{tabular}

}
\end{center}
\vskip 0.6cm

Keeping all that in mind, we performed Monte Carlo analysis of jet production
 and found the variable that satisfied all
the criteria. The detailed description is given below and started with definitions
of variables used in jet production
analysis.

{\subsection{Main definitions}}
Jets are experimentally defined as the amount of energy deposited in the cone of radius
$R=\sqrt {(\Delta \eta)^2 +{(\Delta \phi)}^2 }$ in the space
$(\eta, \phi)$, where
$\Delta \eta$  and $\Delta \phi$ specify the extent  of the cone
 in the pseudorapidity  and azimuth.  The pseudorapidity
$\eta$ is determined via the center mass angle $\theta$
by the formula $\eta=-ln(tg(\theta/2))$. In this work, the value of cone radius was taken
to be $R=0.7$.

The inclusive jet cross section measures the probability of observing a hadronic jet with
a given $E_T$ and $\eta$ in a hadron-hadron collision.
The inclusive jet cross section is usually expressed in terms of the invariant cross section
\begin{equation}
E\frac{d^3\sigma}{dp^3}.
\label{eq:r1}
\end{equation}
In the experiments \cite{CDF, D0}, the measured variables are the transverse energy ($E_T$)
and pseudorapidity
($\eta$). In terms of these variables, the cross section is expressed as follows
\begin{equation}
\frac{d^2\sigma}{d{E_T}d{\eta}}.
\label{eq:r2}
\end{equation}
The quantities (\ref{eq:r1}) and (\ref{eq:r2}) are related by
\begin{equation}
E\frac{d^3\sigma}{dp^3} = \frac{1}{2\pi E_T}\frac{d^2\sigma}{d{E_T}d{\eta}}.
\label{eq:r3}
\end{equation}
The expression (\ref{eq:r3})    follows if the jets are assumed
to be massless. For most measurements, the cross section is averaged over some range of
pseudorapidity. In this paper as in \cite{CDF, D0}, we analyzed jets in central pseudorapidity
region $ |\eta |< 0.5$.

{\subsection{Results}}
The PYTHIA6.2 Monte Carlo generator \cite{PYTHIA} is used for
calculation of inclusive jet cross sections in hadron-hadron
($pp$, ${\bar p}p$)
collisions in high energy range and for pseudorapidity $\eta = 0$.
As a first step, we simulated the inclusive jet cross sections at Tevatron energies
$\sqrt s$= 1800 and 630~GeV.
The comparison between the Monte Carlo simulations and experimental data \cite{D0} is shown in Figure 2.
Black points denote experimental data while red crosses denote Monte Carlo results.
The agreement is very good.
But, this agreement can be obtained
only if the higher-order effects are included in PYTHIA code. This can be
done in PYTHIA 6.2 by including the so-called K-factor. K-factor is the
ratio of NLO cross section and LO cross section. In this case, we used the
model with separate factors for ordinary and colour annihilation graphs.
As expected, we can see very strong dependence on the jet transverse energy
$E_{T}$.

In order to compare jets cross sections at two different colliding energies
the so-called "scaled dimensionless cross section" (SDCS)
is used. This variable reads
\begin{equation}
SDCS = E^4_T \cdot E\frac{d^3\sigma}{dp^3}.
\end{equation}
The scaling hypothesis, which is motivated by the
Quark-Parton Model, predicts that this variable plotted against $x_T = (2E_T/\sqrt{s})$
will be independent of the collision energy $\sqrt s $.
However, QCD leads to scaling violation through the running coupling constant
$\alpha_s$ and the
evolution of the PDF's \cite{D0}.
Theoretically, the scaled dimensionless cross sections at different collision energies
should be nearly exponential and close to one another \cite{D0}, or in other words,
the ratio of CDCS's for different energies
should be a constant
when plotted as a function of $x_T$. Figure 3(a) shows the ratio of dimensionless
inclusive jet cross sections at $\sqrt s$ = 630 and 1800 GeV and for $ |\eta |< 0.5$
as well as corresponding results of Monte
Carlo simulations\footnote{J. Womersley wrote about these results: "Both CDF and DO
 have measured the ratio of
jet cross sections, exploiting a short period of data taking at the latter center
of mass energy at the end of Run I. This ratio is expected to be a rather straightforward
quantity to measure and to calculate. Unfortunately, the two experiments
are not obviously consistent with each other (especially at low $x_T$ ) or with the NLO
QCD expectation for the ratio. At least two explanations have been
suggested for the discrepancy. It seems that more work, both theoretical and
experimental, is needed before this question can be resolved."}.

This variable was our starting point, because
it practically does not depend on $x_T$. It can be seen in Figure 3(a)  that
$SDCS(630) / SDCS (1800) > 1$. It means that the SDCS decreases with increasing colliding energy.
On the other hand, the scale factor for jets $\rho_{jet}$ should take into account the rise
of the jet $E_T$ with increasing $\sqrt s$ (for the same $x_T$ bin) and the behavior of
$\sigma_{jet}$ for minijets production which increase with energy approximately as
$s^\delta  ln(s)$, where $\delta$ value is between 0 and 0.4 (from QCD expectations
and HERA results).

So, the next step was to find the variable similar to the SDCS with taking into account above
requirements. The natural choice was to multiply the SDCS value with
corresponding jet $E_T$ and to divide with number of jets in $\eta$ region. The
variable, we called "scaled jet energy density", then has the form:
\begin{equation}
Scaled~jet~energy~density = SDCS \cdot \frac{E_T}{N_{jet}}.
\end{equation}


The shape of this variable for different colliding energies is $x_T$
independent. It increases with energy $\sqrt s $ (from 200 to 14000~GeV).
The results are shown in Figure 3(b).
This variable has the straightforward interpretation - it reflects
the probability to have a jets
with defined $E_T$ in defined $x_T$ and pseudorapidity region
(in this case we talk about central region).
The $x_T$ independence of scaled jet energy density is clearly
 seen when the values are normalized.
Following the procedure described in \cite{TD} we applied
the normalization by dividing the scaled jet
energy density values for different $\sqrt s$ with corresponding
value for $\sqrt s$ = 1800~GeV.
The results are given in Figure 4(a). These ratios show one
interesting property: they do not depend on the value
of cone radius $R$ in jet finder algorithm. This is important
 feature because there is a indication that jets at LHC
will be broader than expected \cite {Atlas}.
So, on the basis of these variable properties, we conclude that
this variable can be a good candidate for the role
of scale factor $(\rho_{jet})$ for jet analysis in the framework of $z$-presentation.

Thus,  Figure 4(b) shows the ratio $\rho_{jet} (\sqrt s) /\rho_{jet}
(1800)$ as a function of $\sqrt s$ compared with predictions of $z$-scaling. The first
impression is that corresponding values agree very well up to Tevatron energies.  In
other words, introduction of new variable fully confirms $z$-scaling predictions for jets for
available energies.

{\section{Discussion}}
However, we should be very careful with results of Monte Carlo simulations at
LHC energy. Extrapolations to LHC energies, based on measurements at the Tevatron show the
importance of taking
into account the processes when (relatively) small transverse momenta are involved.
The description of this problem is given in \cite{Mor}: "Most of the time the protons
 will pass through each other
with low amount of momentum (low-$p_T$) being transferred between the interacting partons.
Occasionally there will be a "hard" parton-parton collision, resulting in large transverse
momentum outgoing particles.
Perturbative QCD is highly successful when applied to hard processes (large-$p_T$) but
cannot be applied to soft interactions (low-$p_T$). Alternative approaches to describe soft
processes are therefore required. PYTHIA's model for hadron-hadron collisions
attempts to extend perturbative (high-$p_T$) picture down to low-$p_T$
region considering the possibility that multiple parton scattering takes
place in hadron-hadron collisions". These "soft" processes can violate expected distribution
even in hard processes.
For example, the violation of KNO scaling is also attributed to secondary processes taking
place in the hadron scattering.

The problem of accounting low-$p_T$ processes is present at the Tevatron energies, too.
It was found that the default PYTHIA settings does not describe the minimum bias and underlying event data
at CDF and D0 experiments.
But, with appropriate tunings for PYTHIA \cite{Mor, Kor} those minimum bias and underlying
event data can be described.
So-called CDF - tune A is the best model describing experimental data from
the Tevatron. However, it fails to reproduce several minimum bias distributions at lower
energies. On the other hand,
tune from \cite{Mor} gives reasonable description of underlying event data
 and nice description of minimum bias distributions.
 The relevant PYTHIA6.2 parameters values in different
tuning \cite{Mor, Kor} are shown in Table 2.

This problem is interesting also in the case of jet production.
It should be noted that the results for jet energy density shown in Figures 2-4
are obtained with CDF Tune A
parameters with fixed $p_T$ cut for multiple interactions at different collision energies.
The changes in results compared with default PYTHIA values are small at energies up to 1800~GeV,
but situation could be quite different at the LHC.

\begin{center}
\vskip 0.4cm
\noindent
{\it{Table 2.} The relevant PYTHIA6.2 parameters values in different tuning \cite{Mor, Kor}.}

\vskip 0.5cm
{\baselineskip=1.5\baselineskip
\begin{tabular}{|c|c|c|c|c|}
\hline
{Parameter} & {Deafult} & {CDF - Tune A} & {Moraes Tune} & {Our Tune}\\ \hline
{PARP(67)} & {1.0} & {4.0} & {1.0} & {1.0}\\ \hline
{MSTP(82)} & {1} & {4} & {4} & {4}\\ \hline
{PARP(82)} & {1.9} & {2.0} & {1.8} & {1.8}\\ \hline
{PARP(84)} & {0.2} & {0.4} & {0.5} & {0.6}\\ \hline
{PARP(85)} & {0.33} & {0.9} & {0.33} & {0.66}\\ \hline
{PARP(86)} & {0.66} & {0.95} & {0.66} & {0.66}\\ \hline
{PARP(89)} & {1000} & {1800} & {1000} & {1000}\\ \hline
{PARP(90)} & {0.16} & {0.25} & {0.16} & {0.16}\\ \hline
\end{tabular}

}
\end{center}

At the LHC the important topic will be multiple parton scattering
i.e. the simultaneous occurrence of two independent hard (semihard, soft) scattering
in the same interaction.
On the other hand, in a hard scattering process,
the underlying event has a hard
component (initial + final-state radiation and particles from the
outgoing hard scattered partons) and a soft component (beam-beam
remnants). In case of such extreme colliding energies the small differences
 in "physical scenarios" can produce sizeable
differences in scaled jet energy density. Analyzing the parameters values
in Table 2, all the tunes assume
smooth transition between high and low-$p_T$ regions ($MSTP(82)=4$) instead
of cut on $p_T$ ($MSTP(82)=1$).
It can be seen that main changes are for values of parameters PARP(84), PARP(85)
and PARP(86). PARP(84) regulates the size
of the hadron core if the double Gaussian matter distribution in hadrons is assumed.
PARP(85) and PARP(86) describe the
probability that multiple parton scattering produces two gluons with color connections
to the nearest neighbors or as a closed gluon loop.
We also applied our tune by increasing the probability of producing two gluons with color
 connections to the nearest neighbors
in multiple interactions and by increasing the size of of the hadrons core
(right column in Table 2). This results in decrease
of scaled jet energy density ratio at the LHC energies and
corresponding values are very close to prediction of Z-scaling.
Comparing the values of jet energy density at the LHC energy (Figure 4(b)) simulated with
different tuning \cite{Mor, Kor} and our tune, it can be concluded that this variable is sensitive
(at the level of $10 \div 20\%$) to the changes of these parameters.

\vspace*{-0.5cm}

{\section{Conclusions}}
In this work we tried to find the variable which will reflect
the main features of the jet production in low transverse energy range at a given energy
and play the role of the scale factor for description of jets in the framework of $z$-scaling.
The PYTHIA6.2 Monte Carlo generator was used for calculation of jet
production in proton-proton collisions over a
high energy range ($\sqrt s = 100 \div 14$~TeV) and for pseudorapidity $\eta = 0$.
We introduced the
variable we called the "scaled jet energy density".
The scaled jet energy density is the probability to have a jet
with defined $E_T$ in defined $x_T$ and pseudorapidity regions.
Its definition is related to the "scaled dimensionless cross section" and its features
(for example, $x_T$ independence) show that this variable can be used
in the studies of jet production at high energies.
The important result is that properties of new variable fully confirms $z$-scaling predictions
for jets production at available energies.
Detailed analysis of the variable behavior at the LHC energies show that it is sensitive to
relatively small differences in applied "physical scenarios" in standard Monte Carlo generators.
The fact is that there are sizeable uncertainties in LHC predictions
generated by different models so
the alternative approach as $z$-scaling is very important for understanding
of inclusive processes of jet production at high energies.

\vskip 5mm
{\large \bf Acknowledgments.}
One of the authors (M.T.) would like to thank I.Zborovsk\'{y} for
numerous fruitful and stimulating discussions of the obtained results.


\newpage
\vspace*{3.5cm}
\begin{center}
\hspace*{-3.5cm}
\parbox{5cm}{\epsfxsize=5.cm \epsfysize=5.cm \epsfbox[5 5 350 350]
{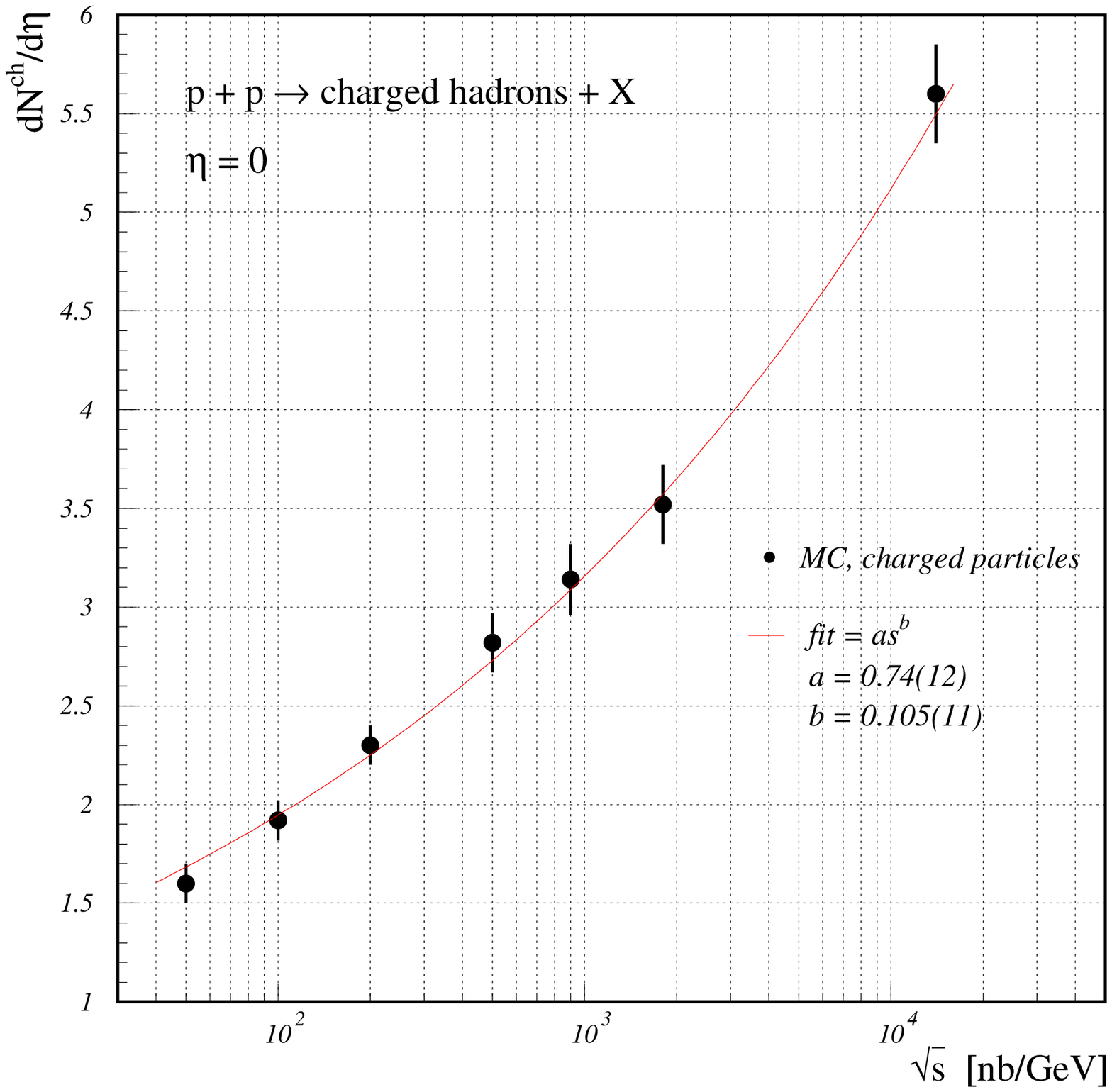}{}} \hspace*{3cm}
\parbox{5cm}{\epsfxsize=5.cm \epsfysize=5.cm \epsfbox[5 5 350 350]
{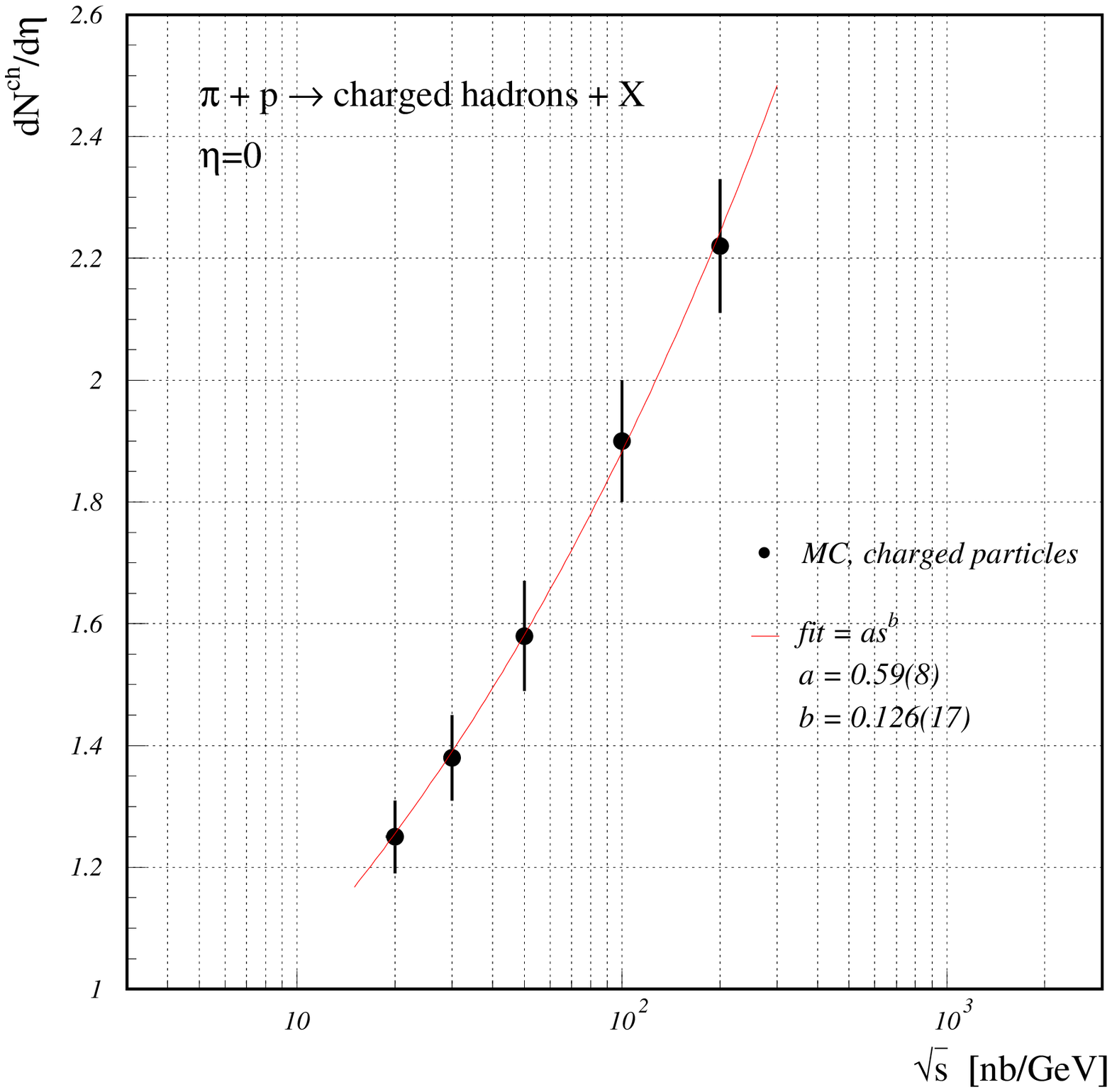}{}} \vskip -1.9cm
\hspace*{-0.4 cm} (a) \hspace*{7.6cm} (b)\\[0.5cm]

\end{center}
\vskip 0.cm

{\bf Figure 1.} (a) Charged particle multiplicity density
$dN^{ch}/d\eta$ at $\eta =0$  as a function of energy $\sqrt s$ in
$pp$ collisions.
 (b) Charged particle multiplicity density $dN^{ch}/d\eta$ at $\eta =0$  as
a function of energy $\sqrt s$ in ${\pi}p$ collisions.

\vskip 6.5cm

\begin{center}
\hspace*{-3.5cm}
\parbox{5cm}{\epsfxsize=5.cm \epsfysize=5.cm \epsfbox[5 5 350 350]
{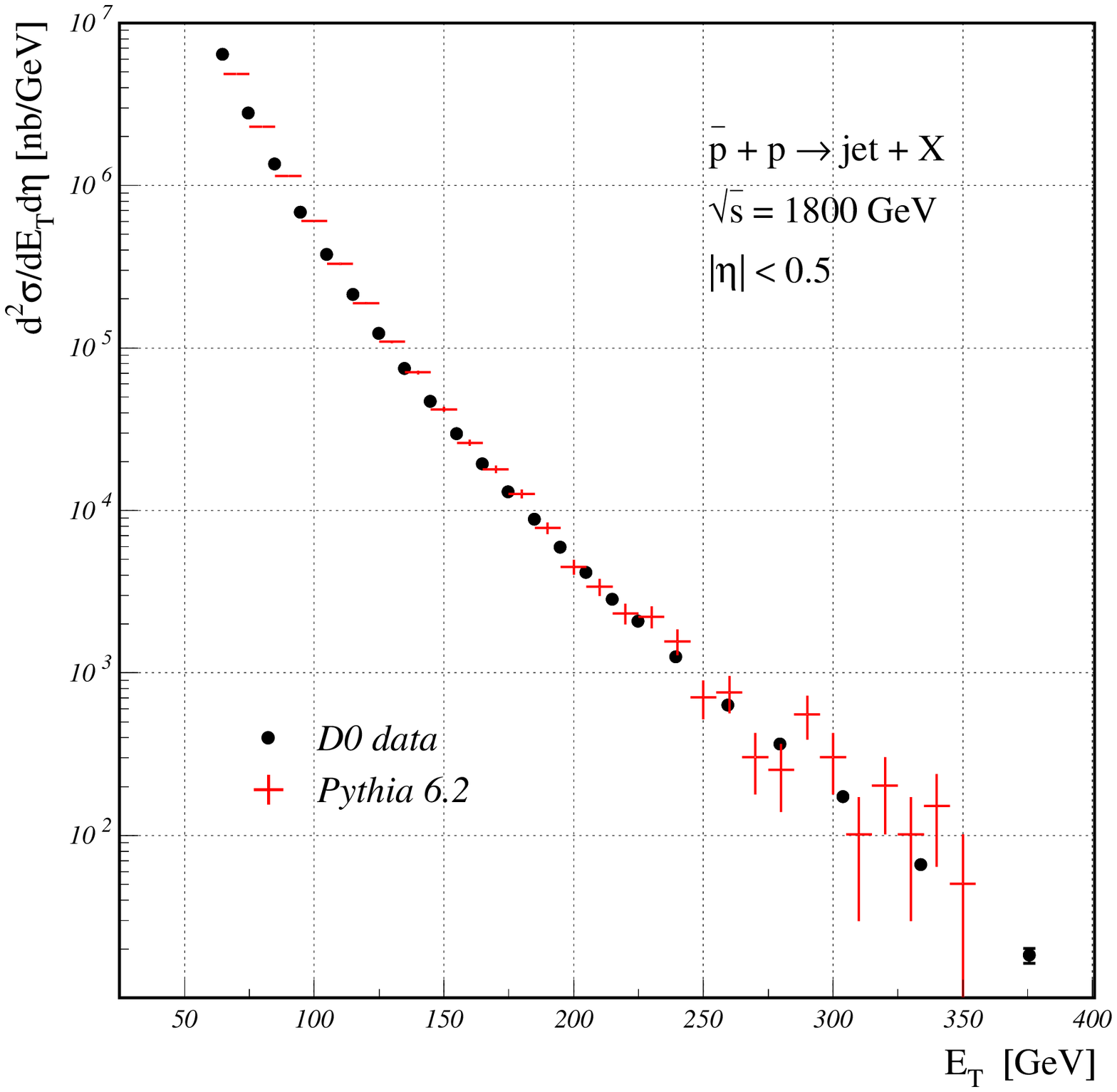}{}} \hspace*{3cm}
\parbox{5cm}{\epsfxsize=5.cm \epsfysize=5.cm \epsfbox[5 5 350 350]
{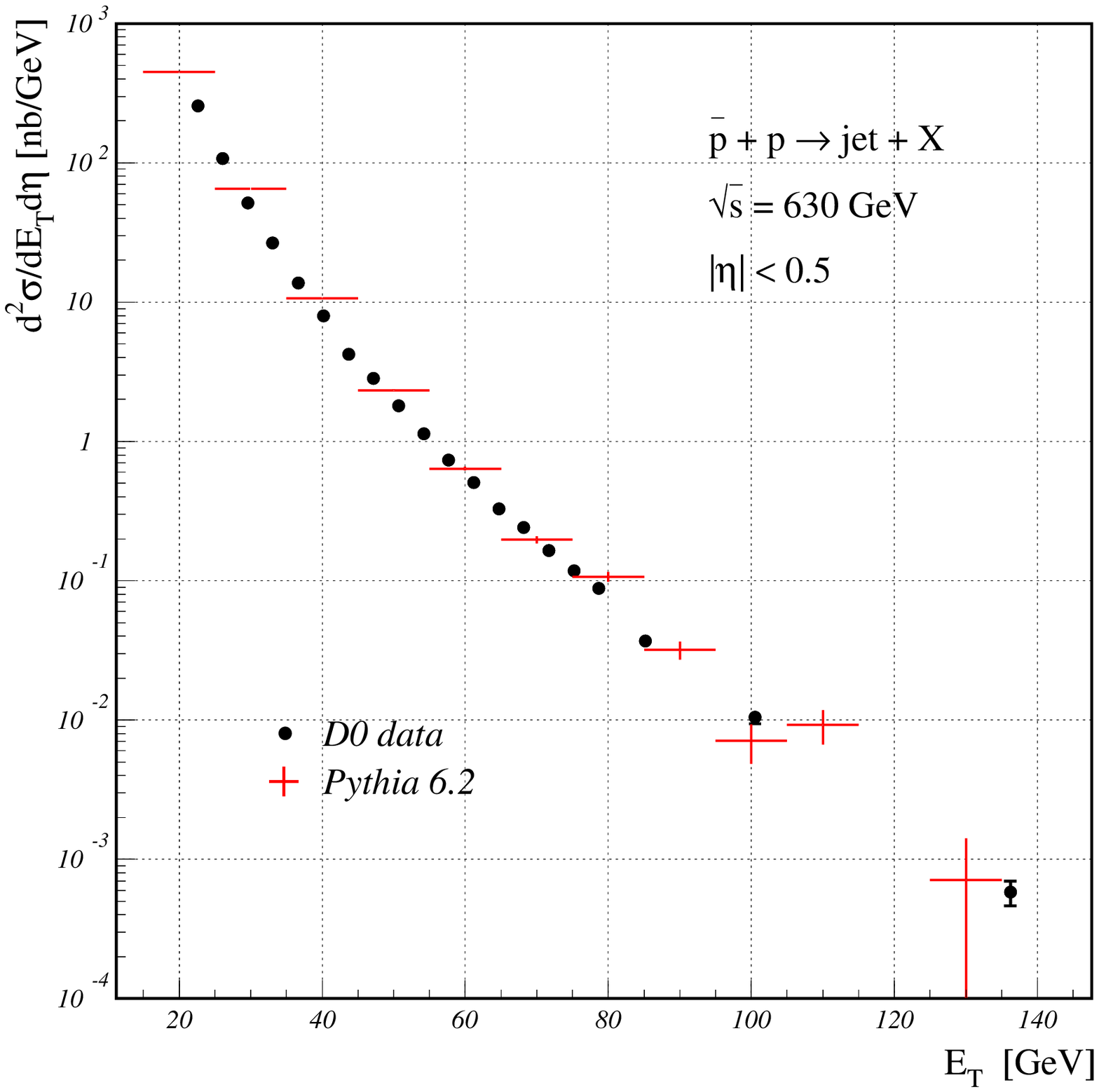}{}} \vskip -1.9cm
\hspace*{-0.4 cm} (a) \hspace*{7.6cm} (b)\\[0.5cm]

\end{center}
\vskip 0.cm

{\bf Figure 2.} The comparison between the MC simulations and
experimental data on inclusive jet cross sections for $ |\eta |<
0.5$ in $\bar pp$ collisions at Tevatron energies $\sqrt s$= 1800
and 630~GeV \cite{D0}. Black points - experiment, red crosses -
Monte Carlo results.


\newpage
\vspace*{3.5cm}
\begin{center}
\hspace*{-3.5cm}
\parbox{5cm}{\epsfxsize=5.cm \epsfysize=5.cm \epsfbox[5 5 350 350]
{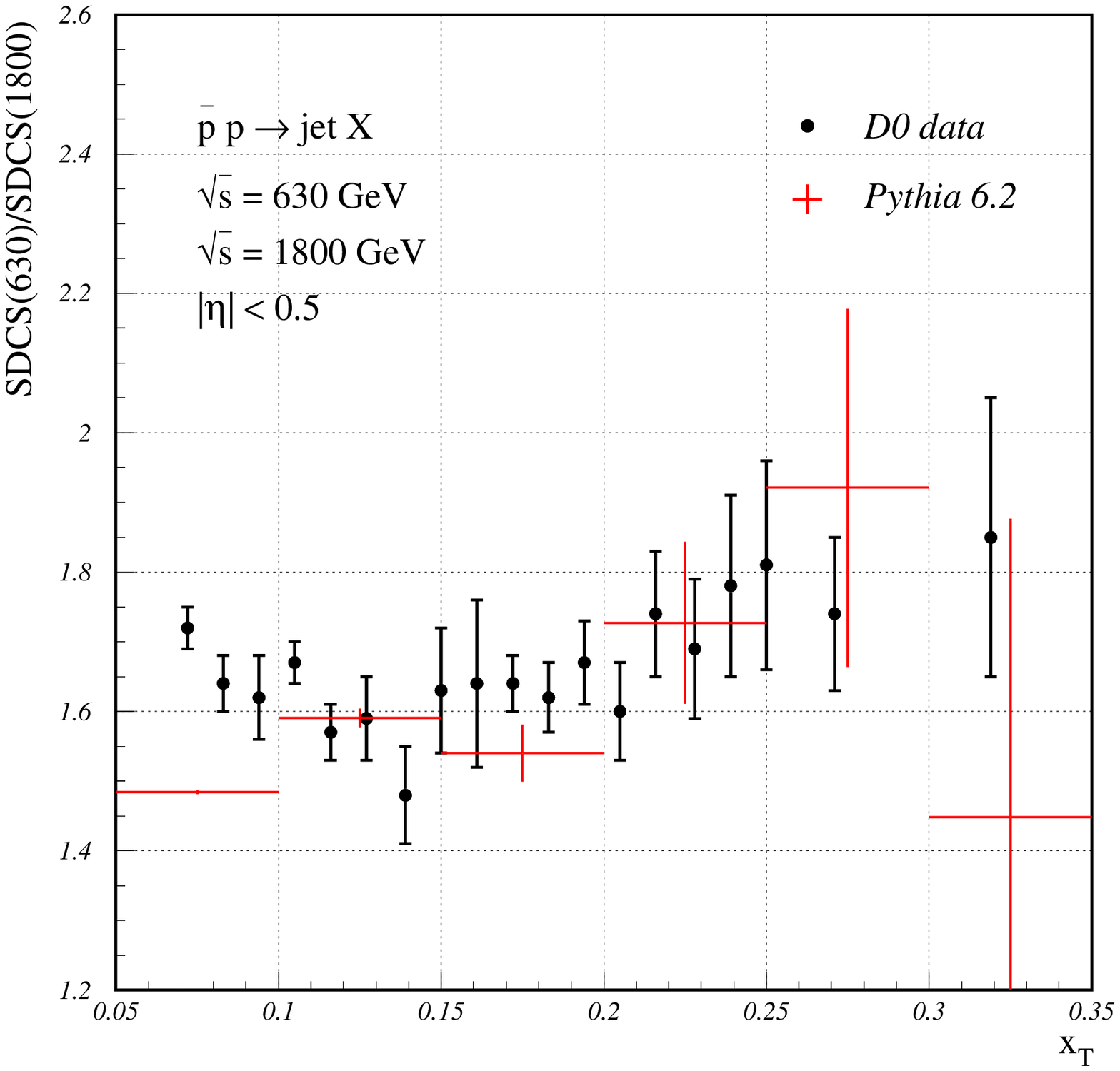}{}} \hspace*{3cm}
\parbox{5cm}{\epsfxsize=5.cm \epsfysize=5.cm \epsfbox[5 5 350 350]
{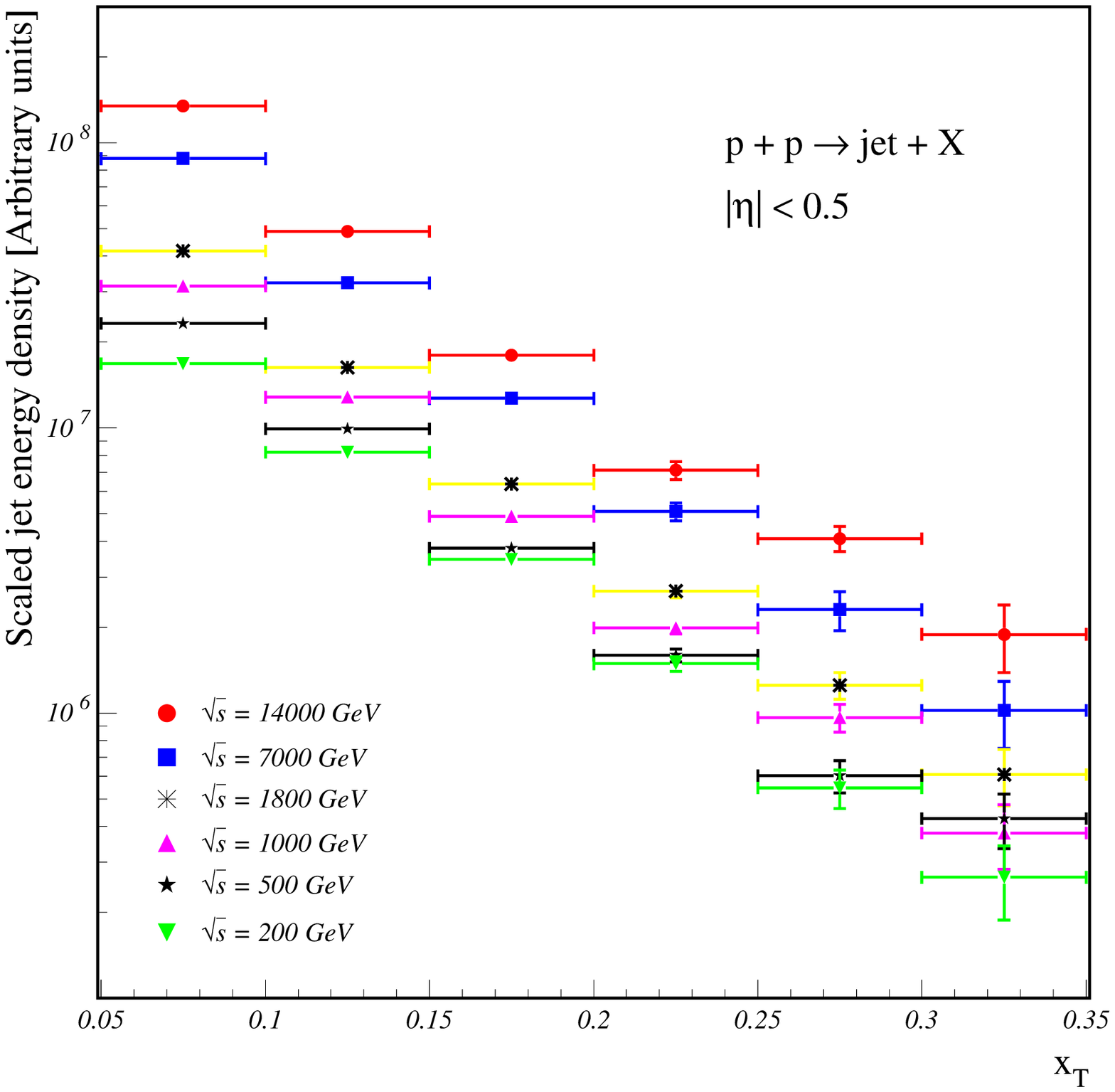}{}} \vskip -1.9cm
\hspace*{-0.4 cm} (a) \hspace*{7.6cm} (b)\\[0.5cm]

\end{center}
\vskip 0.cm

{\bf Figure 3.}  (a) The ratio of dimensionless inclusive jet
cross sections at $\sqrt s$ = 630 and 1800~GeV and for  $ |\eta |<
0.5$ in comparison with corresponding results of Monte Carlo
simulations.
 (b) The scaled jet energy density in central pseudorapidity
 region for different collision energies (from 200 to 14000~GeV)
as a function of $x_T$.

\vskip 6.5cm

\begin{center}
\hspace*{-3.5cm}
\parbox{5cm}{\epsfxsize=5.cm \epsfysize=5.cm \epsfbox[5 5 350 350]
{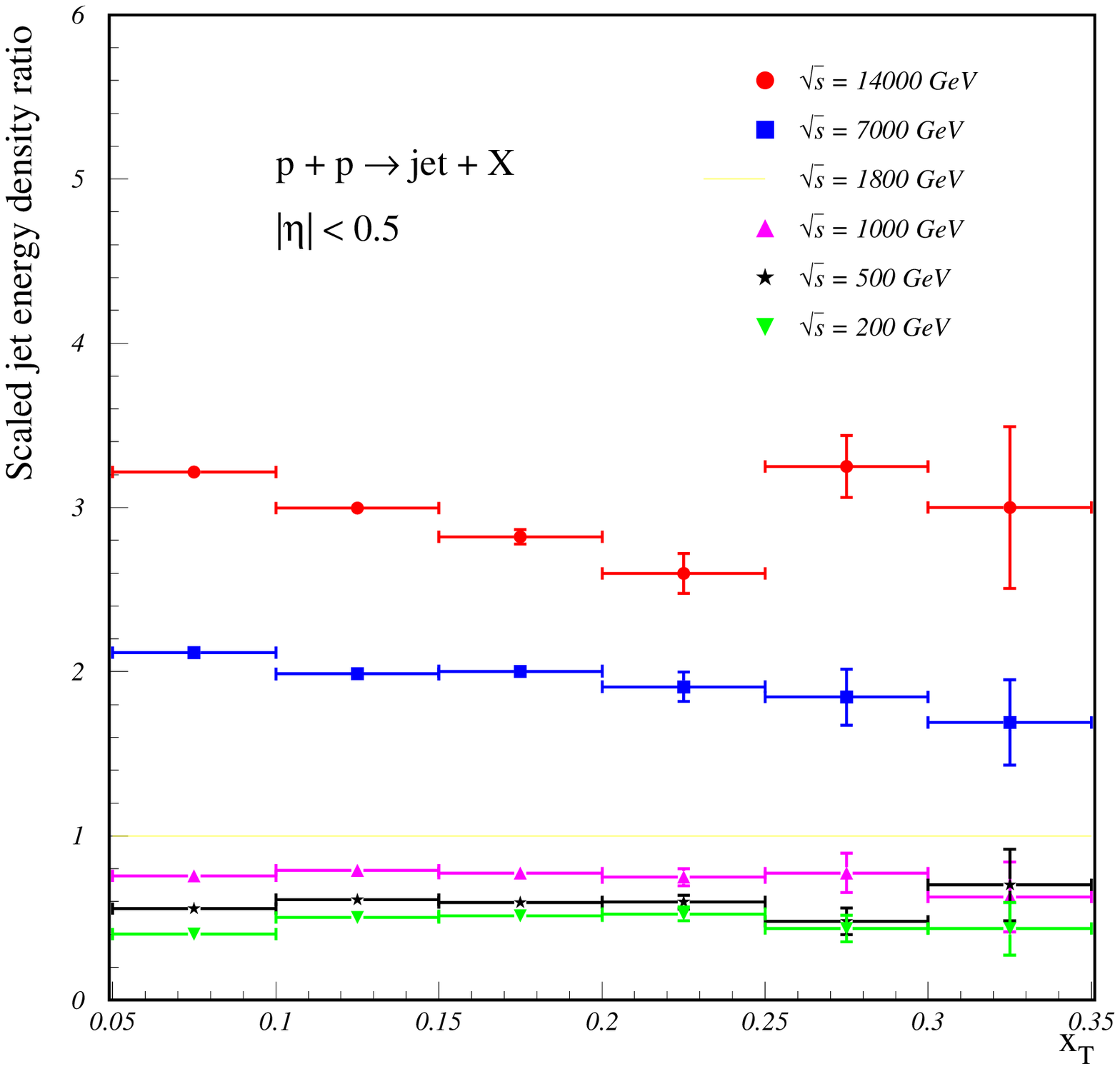}{}} \hspace*{3cm}
\parbox{5cm}{\epsfxsize=5.cm \epsfysize=5.cm \epsfbox[5 5 350 350]
{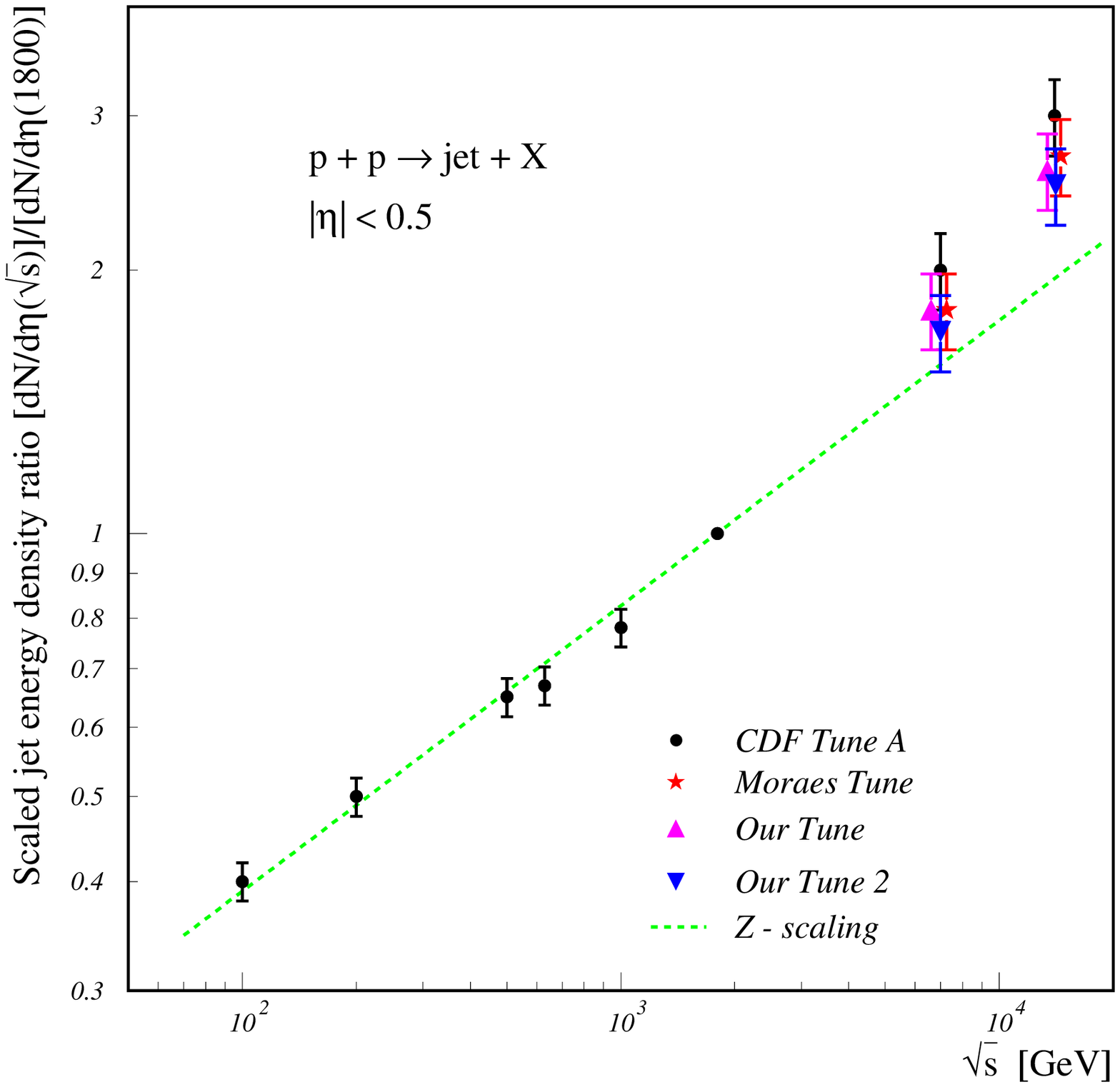}{}} \vskip -1.9cm
\hspace*{-0.4 cm} (a) \hspace*{7.6cm} (b)\\[0.5cm]

\end{center}
\vskip 0.cm

{\bf Figure 4.}
 The scaled jet energy density ratio (see text) in central
pseudorapidity region for different collision energies (from 200
to 14000~GeV) as a function of $x_T$ (a) and  $\sqrt s$ (b).

\end{document}